# Electron Localization in Non-Compact Covalent Bonds Captured by the r$^2$SCAN+$V$ Approach

Yubo Zhang[1,*], Da Ke[1], Rohan Maniar[3], Timo Lebeda[3], Peihong Zhang[2,*], Jianwei Sun[3,*], John P. Perdew[3,*]

[1]*Minjiang Collaborative Center for Theoretical Physics, College of Physics and Electronic Information Engineering, Minjiang University, Fuzhou, China*

[2]*University at Buffalo, State University of New York, Buffalo, New York 14260, USA*

[3]*Department of Physics and Engineering Physics, Tulane University, New Orleans, Louisiana 70118, USA*

Corresponding emails: yubo.drzhang@mju.edu.cn, pzhang3@buffalo.edu, jsun@tulane.edu, perdew@tulane.edu

**Abstract:** In density functional theory, the SCAN (Strongly Constrained and Appropriately Normed) and r$^2$SCAN functionals significantly improve over generalized gradient approximation functionals such as PBE (Perdew–Burke–Ernzerhof) in predicting electronic, magnetic, and structural properties across various materials, including transition-metal compounds. However, there remain puzzling cases where SCAN/r$^2$SCAN underperform, such as in calculating the band structure of graphene, the magnetic moment of Fe, the potential energy curve of the Cr$_2$ molecule, and the bond length of VO$_2$. This research identifies a common characteristic among these challenging materials: non-compact covalent bonding through *s-s*, *p-p*, or *d-d* electron hybridization. While SCAN/r$^2$SCAN excel at capturing electron localization at local atomic sites, they struggle to accurately describe electron localization in non-compact covalent bonds, resulting in a biased improvement. To address this issue, we propose the r$^2$SCAN+$V$ approach as a practical modification that improves accuracy across all the tested materials. The parameter $V$ is 4 eV for metallic Fe, but substantially lower for the other cases. Our findings provide valuable insights for the future development of advanced functionals.

**Significance statement:**

To predict material properties, accurate but efficient approximations for the electronic exchange-correlation energy are needed. r$^2$SCAN satisfies many more exact conditions than PBE, and is more accurate in most cases, but there are puzzling cases where PBE is more accurate, including two-dimensional carbon, metallic iron, the chromium dimer, and solid vanadium dioxide. We show that these errors of r$^2$SCAN can be strongly reduced by making a one-parameter +$V$ correction that localizes electrons near bond centers between nuclei, suggesting a correction of self-interaction error in non-compact covalent bonds. In such situations, r$^2$SCAN may provide much of the +$U$ correction to PBE that localizes electrons near nuclei, and not so much of the compensating +$V$.

# 1. Introduction

Kohn-Sham density functional theory (DFT) [1] with the generalized gradient approximation (GGA) in the form of Perdew-Burke-Ernzerhof (PBE) [2] has long been the workhorse of condensed matter physics and materials science, but its reliability degrades for transition-metal materials. For example, PBE frequently fails to capture the insulating characteristics of some Mott insulators. PBE's primary limitation lies in its inadequate treatment of electron localization, giving rise to the self-interaction error (SIE), delocalization error, and strong correlation error [3,4,5]. A popular remedy is introducing an on-site Hubbard-$U$-like potential, expressed as $\Delta P_i = P_{\text{DFT}+U} - P_{\text{DFT}} = U(\frac{1}{2} - n_i)$ [6,7]. $\Delta P_i$ is negative for filled orbitals (with occupation $n_i = 1$) and positive for empty ones ($n_i = 0$). This corrective potential increases the energy splitting between occupied and unoccupied states, which explains its effectiveness in opening bandgaps of Mott insulators. In addition, the $U$ potential enhances a site-localized orbital's spatial localization by reducing SIE [8].

Although the $U$ value can be determined from first principles [8], it is more commonly treated as an empirical parameter to reproduce some experimental quantities. Additionally, it has been observed that the GGA+$U$ method can inadvertently suppress covalent hybridization between localized $d$ orbitals and more dispersive $p$ orbitals of anions [9,10]. For instance, in multiferroic $YMnO_3$, a minimum $U$ value of 5 eV is required to open a bandgap; but this value is so large that it nearly suppresses $p$-$d$ hybridization, rendering the $d$ orbitals irrelevant for ferroelectric-ferromagnetic coupling [10].

Hybrid functionals, such as the Heyd-Scuseria-Ernzerhof (HSE) functional [11,12], demonstrate superior predictive accuracy for many semiconductors and insulators. However, HSE faces accuracy challenges when applied to transition-metal materials, particularly metallic systems [13]. For instance, HSE06 incorrectly opens a bandgap in metallic $La_{2-x}Sr_xCuO_4$ [14], making this functional unsuitable for studying metal-insulator transitions. Additionally, the computational cost is prohibitively high for calculating large systems.

Introduced in 2015, the *Strongly Constrained and Appropriately Normed* (SCAN) meta-GGA functional [15] represents a remarkable advancement in DFT. Later, the r$^2$SCAN [16] was developed to improve the numerical efficiency. As general-purpose functionals, SCAN/r$^2$SCAN have demonstrated high accuracy across a broad spectrum of materials, including transition-metal compounds [17,18,19]. For example, in cuprate materials, SCAN not only accurately captures the pristine insulator and hole-doped metal [20], but also reliably predicts the emergence of a striped phase characterized by the coexistence of insulating and metallic regions [21]. The improved performance is primarily attributed to SCAN/r$^2$SCAN's ability to mitigate SIE [22,23]. Additionally, SCAN/r$^2$SCAN are able to recognize chemical bonds [24]—a critical feature for describing the pronounced anisotropy of $d$ orbitals [10,23]. For instance, in $YMnO_3$, SCAN effectively characterizes various anisotropic orbitals, including highly localized non-bonding $d$ orbitals, less localized bonding $d$ orbitals, and dispersive $sp$ orbitals [10]. These synergistic enhancements enable SCAN to open the bandgap of $YMnO_3$ without distorting the $p$-$d$ hybridization. For many but not all transition metal compounds, SCAN and r$^2$SCAN can still be improved by +$U$ corrections, but the needed $U$ values are significantly smaller than for PBE [18]. Significant improvements without +$U$ in the progression from local spin-density approximation [1] to the PBE GGA to the r$^2$SCAN meta-GGA have been found for the first three ionization energies of the 3$d$ atoms [25], and for the oxidation energies of the ionically-bonded 3$d$ transition-metal oxide solids [26].

SCAN and r$^2$SCAN also well capture the strong covalent bond energetics found in many main-group molecules at equilibrium [15,16], where strong hybridization of dispersive $s$ and $p$ orbitals can produce relatively *compact* bond orbital shapes. However, weak or stretched bonds and bonds involving localized $d$ electrons are more prone to SIE in these functionals. A canonical example is the $H_2$ molecule, where delocalized *s* orbitals form a *compact* bond near the equilibrium position but gradually lose their compactness as the bond is stretched, due to the increasing orbital localization. We observe that r$^2$SCAN's accuracy diminishes as the bond transitions from the *compact* to the *non-compact* region (see Figure S6 in the Supplementary Materials).

Despite SCAN/r$^2$SCAN's overall success, there are puzzling instances where they only match or underperform compared with PBE. A prominent example is SCAN's overestimation of the magnetic moment in elemental iron [27], predicting a value of 2.75 $\mu_B$, which exceeds the experimentally measured 2.22 $\mu_B$. In comparison, PBE offers a more accurate estimate of 2.20 $\mu_B$. Additionally, SCAN faces accuracy issues in predicting the potential energy curve of the $Cr_2$ molecule, showing greater deviations than PBE [28]. SCAN/r$^2$SCAN also incorrectly open a bandgap for graphene [28] and inaccurately calculate the bond length of $VO_2$ [29].

Shifting perspectives, it is intriguing that PBE outperforms SCAN/r$^2$SCAN in these materials, despite PBE's more pronounced SIE. This observation mirrors a similar finding in $BaTiO_3$, where the oldest local spin-density approximation (LSDA) reproduces the ferroelectric polarization with an unexpected accuracy because of a fortuitous error cancellation [30]. LSDA underestimates the geometric distortion while overestimating electronic polarizability. In contrast, PBE simultaneously overestimates both quantities, leading to a significant deviation in the calculated ferroelectric polarization. By comparison, SCAN accurately predicts the ferroelectricity for the correct reasons.

This work reveals that Fe, $Cr_2$, graphene, and $VO_2$ share a common bonding characteristic: non-compact covalent bonding through *s-s*, *p-p*, or *d-d* electron hybridization, which can lead to notable electron localization between atomic sites. While r$^2$SCAN improves upon PBE in capturing electron localization at the *local atomic sites*, it fails to adequately describe the electron localization in the *non-compact bonds*. PBE's superior performance arises from a coincidental cancellation of errors, despite suffering from more pronounced SIE in both regions. To address this issue, we propose augmenting r$^2$SCAN with a corrective potential *V*, which systematically improves the calculations across all tested materials. Practically, the r$^2$SCAN+*V* approach is more manageable than the original GGA+*U*+*V* approach [31,32,33,34,35] for difficult materials, as demonstrated in this work.

For computational and other details, see the Supplementary Materials.

## 2. Graphene with *p-p* hybridization

Graphene is a semimetal characterized by linearly dispersive valence and conduction bands that meet at the Dirac point. When spin-symmetry-breaking is allowed, SCAN/r$^2$SCAN open a bandgap (Figure 1a) by stabilizing local magnetic moments that are antiparallelly coupled on the bipartite lattice [28]. Although this scenario is problematic due to accuracy issues, it represents the intriguing Mott gapping mechanism [36,37]— one of the key efforts to transform graphene into a semiconductor. The *Hubbard model* study has established that Coulomb interactions between $p_z$ orbitals are notably strong, approximating or surpassing the threshold for spontaneous bandgap formation [38]. However, investigations on the *extended Hubbard model* [39,40] reveal that nonlocal inter-site interactions effectively screen these Coulomb interactions, thereby suppressing the bandgap formation [37,38,41].

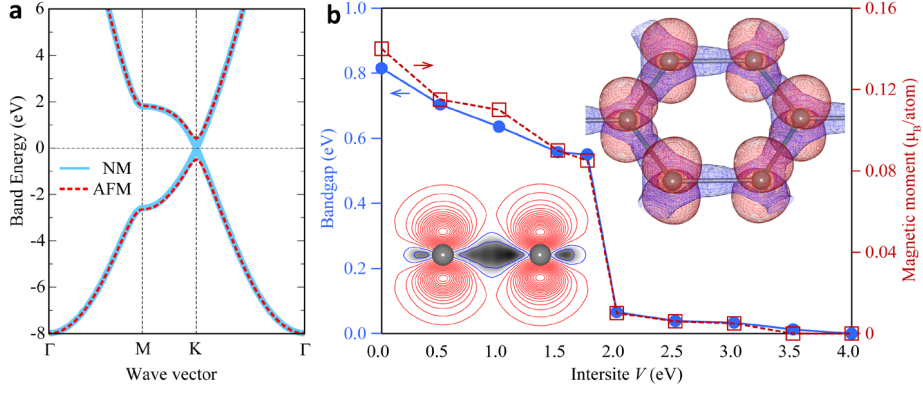

**Figure 1. Electronic properties of graphene.** (a) Bands derived from the $p_z$ orbitals. The calculation uses r²SCAN, with and without spin-symmetry-breaking. (b) Bandgap and magnetic moment as functions of $V$ values in r²SCAN+$V$. The insets illustrate electron redistribution patterns induced by the $V$ (2 eV) potential, represented as $\Delta n = n^{\text{r}^2\text{SCAN}+V} - n^{\text{r}^2\text{SCAN}}$. The upper inset is a 3D plot, while the lower inset provides a 2D cross-section of a C-C bond perpendicular to the lattice plane. Red and blue lines indicate electron depletion and accumulation, respectively.

Drawing upon the wisdom of the *Hubbard model*, the DFT+$U$ approach [6,7] is a well-established method to improve the on-site electron localization. By contrast, DFT+$V$, where $V$ represents an inter-site corrective potential from the *extended Hubbard model*, has received comparatively less attention and warrants a brief introduction:

$$H = t \sum_{\langle i,j \rangle, \sigma} (c_{i,\sigma}^\dagger c_{j,\sigma} + \text{h.c.}) + U \sum_i (n_{i,\uparrow} n_{i,\downarrow}) + V \sum_{\langle i,j \rangle} (n_i n_j)$$

In this Hamiltonian [31,39,40], $t$ is the kinetic energy, $c_{i,\sigma}^\dagger$ is the creation operator of an electron on site $i$ with spin $\sigma$. $U$ is the on-site energy that quantifies the energy penalty for double occupancy at the same atomic site $i$, signifying local electron-electron repulsion. The inter-site $V$, on the other hand, represents the nonlocal interaction between electrons at neighboring sites $i$ and $j$. When implemented into DFT [31], the corrective potential to the $i$-th site is $\Delta P_i = V \sum_{j \in NN(i)} (\frac{1}{2} - n_j)$, where $n_j$ is the occupation number of the $j$-th orbital; NN($i$) is the set of nearest neighbors of site $i$, and more distant neighbors can also be included similarly. Clearly, a positive $V$ discourages electrons simultaneously sitting on nearby atomic sites, effectively encouraging electrons to accumulate elsewhere such as at a non-compact bond center.

When r²SCAN+$V$ is applied (with positive $V$ potentials to the nearest-neighboring $p$ orbitals), the bandgap closes, with the most dramatic change occurring at $V \approx 2.0$ eV (Figure 1b). The magnetic moment exhibits a similar trend, corroborating that the bandgap arises from the Mott mechanism. The insets depict the electron redistribution pattern: the $V$ potential discourages electron occupation in the $p_z$ orbitals, instead encouraging their accumulation near the bond center. It seems that, compared to PBE, r²SCAN includes an implicit +$U$-like self-interaction correction to the on-site $p_z$ orbitals whose effect on the bandgap and spin polarization is cancelled by the explicit +$V$ correction.

## 3. Cr₂ with *d-d* and *s-s* hybridizations

The Cr₂ molecule features two distinct bonding distances in its potential energy curve: a short bond at 1.68 Å and an extended shelf around 2.4~3.0 Å. This unique characteristic makes Cr₂ an exemplary system for evaluating the predictive power of theories. Recent studies have revealed an intriguing puzzle: while many-body wavefunction methods have shown steady improvements over the years [42], the development of DFT seems to stray from the correct path, with predictions gradually worsening as one climbs Jacob's ladder [43] from GGA to meta-GGA to hybrid functionals (Figure 2a).

Despite its significant SIE, how does PBE achieve high accuracy in capturing the potential energy curve? To mitigate SIE in PBE, we adopt PBE+$U_d$ and PBE+$V_{dd}$ (where the subscripts $d$ and $dd$ indicate corrective potentials on $d$ orbitals) to improve the descriptions of the on-site region and inter-site interaction, respectively. However, the revised results become worse, as PBE+$U_d$ excessively weakens the binding while PBE+$V_{dd}$ overly strengthens it. When SIE is addressed simultaneously in both regions using PBE+$U_d$+$V_{dd}$, the good performance is restored, and the description of the short bond is even improved compared to PBE. These findings clearly indicate that +$U$ and +$V$ have opposing effects and must be considered together to improve upon PBE. However, determining precise $U$ and $V$ values is challenging, and the parameterization from the linear response approach is unfortunately problematic for Cr$_2$ (Figure S2). Here, we fix $U$ = 2.0 eV in PBE+$U_d$ and $V$ = 0.8 eV in PBE+$V_{dd}$, while $V$ = 2.2 eV in PBE+$U_d$+$V_{dd}$ is obtained by fine-tuning the $V$ value with a fixed $U$ = 2.0 eV to best reproduce the binding behavior.

With the aid of the corrective methods, we further investigate PBE's enigmatic efficacy. The +$U_d$ method enhances electron localization on the Cr sites while depleting electrons from the bond center (Figure 2c). This redistribution reduces electronic screening of inter-atomic Coulombic repulsion, causing PBE+$U_d$ to predict an underbinding curve. In contrast, the +$V_{dd}$ correction encourages electron accumulation on the bond (Figure 2d), leading to overbinding and reducing inter-atomic Coulombic repulsion. When applying +$U_d$ and +$V_{dd}$ simultaneously, the overall electron redistribution is relatively weak (Figure 2e), which explains the similar performance of PBE+$U_d$+$V_{dd}$ and PBE. Therefore, PBE's high accuracy arises from its incidental balancing of electron delocalization errors in the site and bond regions.

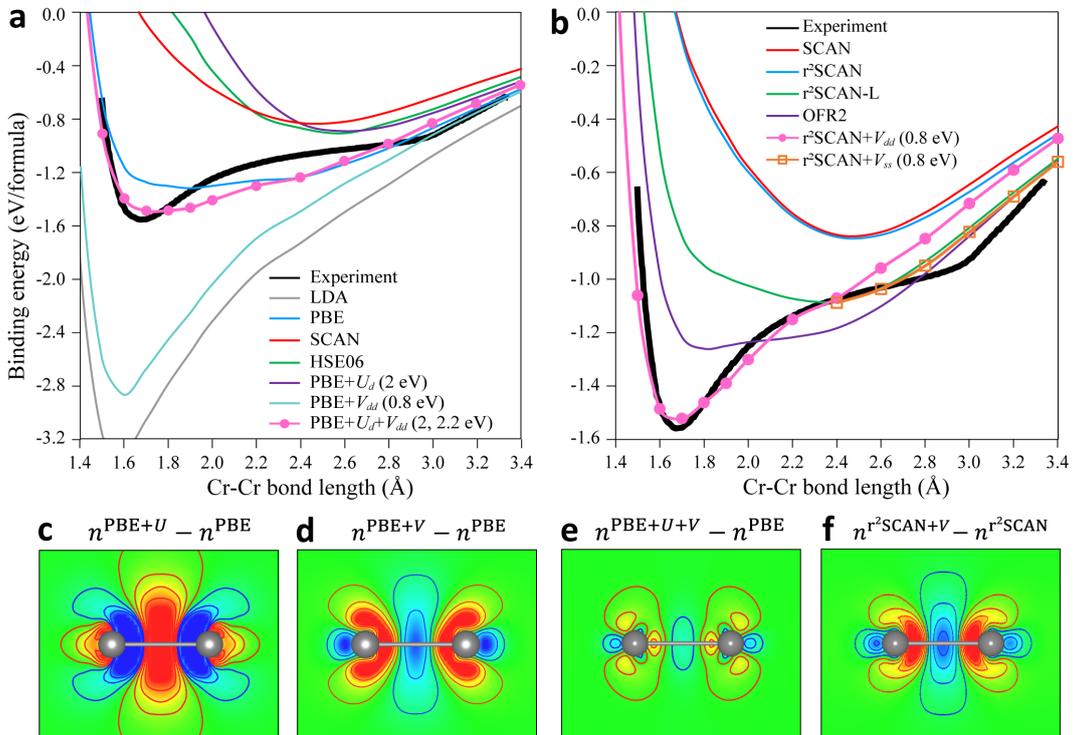

**Figure 2. Electronic properties of the Cr$_2$ molecule.** (a,b) Potential energy curve predicted by various methods. The experimental results are cited from [44]. In subplot (b), while $V_{dd}$ is applied at all bond lengths, $V_{ss}$ is only applied to the shelf structure starting at 2.4 Å. (c) Patterns of electron redistribution due to the additional corrections, represented as $\Delta n = n^{\text{PBE}+U_d(2\text{eV})} - n^{\text{PBE}}$. (d) $\Delta n = n^{\text{PBE}+V_{dd}(0.8\text{eV})} - n^{\text{PBE}}$. (e) $\Delta n = n^{\text{PBE}+U_d(2\text{eV})+V_{dd}(2.2\text{eV})} - n^{\text{PBE}}$. (f) $\Delta n = n^{\text{r}^2\text{SCAN}+V_{dd}(0.8\text{eV})} - n^{\text{r}^2\text{SCAN}}$. Red and blue lines indicate electron depletion and accumulation, respectively. The plots of (c) to (f) correspond to the length of 1.68 Å.

Figure 2b presents the results using meta-GGAs. SCAN/r$^2$SCAN consistently underbind the molecule, similar to PBE+$U_d$, indicating that SCAN/r$^2$SCAN inherently enhance electron localization at atomic sites. Variants like r$^2$SCAN-L [45] and OFR2 [46], which are orbital-free meta-GGAs, show improved performance arising from better accounting of electronic screening effects [46]. However, further enhancements in describing non-compact electron localization are still needed, leading us to propose the r$^2$SCAN+$V$ method. Interestingly, a parameter of $V_{dd}$ = 0.8 eV in r$^2$SCAN+$V$ achieves good agreement with experimental data, particularly in the short bonding region (1.4–2.4 Å). The shelf structure (starting at ~2.4 Å), originating from the hybridization of Cr-4$s$ orbitals [42], is better reproduced when incorporating the intersite potential $V_{ss}$ = 0.8 eV onto neighboring $s$ orbitals. As mentioned in the **Introduction** for H$_2$, $s$-$s$ bonds can become non-compact in stretched configurations, such as in the shelf structure of Cr$_2$. The important role of $V_{ss}$ is further highlighted in the H$_2$ molecule (Figure S6), where the 1$s$ orbital is the sole contributing orbital.

Strictly speaking, the $V$ values should depend on bond lengths due to variations in electron localization. However, r$^2$SCAN+$V$ with the $V_{dd}$ and $V_{ss}$ parameters tightly constrained around 0.8 eV can accurately reproduce the entire curve (Figure S3). The significant improvement in the binding profile, despite the small $V$ parameters, not only highlights the effectiveness of the r$^2$SCAN+$V$ method but also the critical role of functional nonlocality.

Although we identify electron localization in non-compact covalent bonds, we stress that it is not this density change itself but the nonlocality of the functional producing it that is most important for the predicted energetics. We confirmed this for Cr$_2$ by comparing the binding energy curves from self-consistent r$^2$SCAN and from non-self-consistent r$^2$SCAN@r$^2$SCAN+$V$ (r$^2$SCAN evaluated on the r$^2$SCAN+$V$ occupied orbitals and density). The resulting binding energy curves were very similar (Figure S4), as expected, since density-driven errors of the energy [47] tend to be small [48] in the absence of large electron transfers between nuclear basins.

For correlated wavefunction methods, the chromium dimer is a notoriously difficult case of strong correlation [42]. Often symmetry breaking [49,50] transforms strong correlation in a symmetric state into normal correlation in a symmetry-broken state that a reliable density functional can describe, but there is no proof that this must always yield correct energies. In Cr$_2$, the symmetric state is a singlet with zero spin polarization everywhere, and the broken-symmetry state is an antiferromagnetic dimer with a net spin up in one nuclear basin compensated by a net spin down in the other basin.

## 4. VO$_2$ with *d-d* hybridization

VO$_2$, a well-known strongly correlated material, has challenged DFT for decades. Our recent investigation identified a fundamental limitation of the popular functionals: none of LSDA, PBE, r$^2$SCAN, or HSE can accurately describe the vanadium-vanadium dimer length, and incorporating the on-site $U$ correction may even worsen the predictions [29]. Here, we reveal that VO$_2$ exhibits non-compact covalency involving localized $d$ electrons, and r$^2$SCAN+$V$ provides an effective improvement.

Figure 3a compares the dimer length predicted by various methods. PBE underestimates the length (2.53 Å), while r$^2$SCAN (2.66 Å) and HSE (2.70 Å) overestimate it. Incorporating a corrective $U$ (2.0 eV) into PBE, a common strategy for studying VO$_2$, strongly overestimates the bond length (2.83 Å). The +$U$ approach causes an underbinding issue, as revealed by the electron redistribution pattern (Figure 3d): electrons accumulate on the atomic site while being depleted from the bonding region. As a result, the covalent bonding in the dimer is weakened, leading to bond length elongation. Further adding an inter-site $V$ term (i.e., PBE+$U$+$V$) plays a counteracting role by drawing electrons back toward the bond center (Figure 3e). To match the experimental bond length, we fix $U$ at 2.0 eV and identify the optimal $V$ value of 2.2 eV.

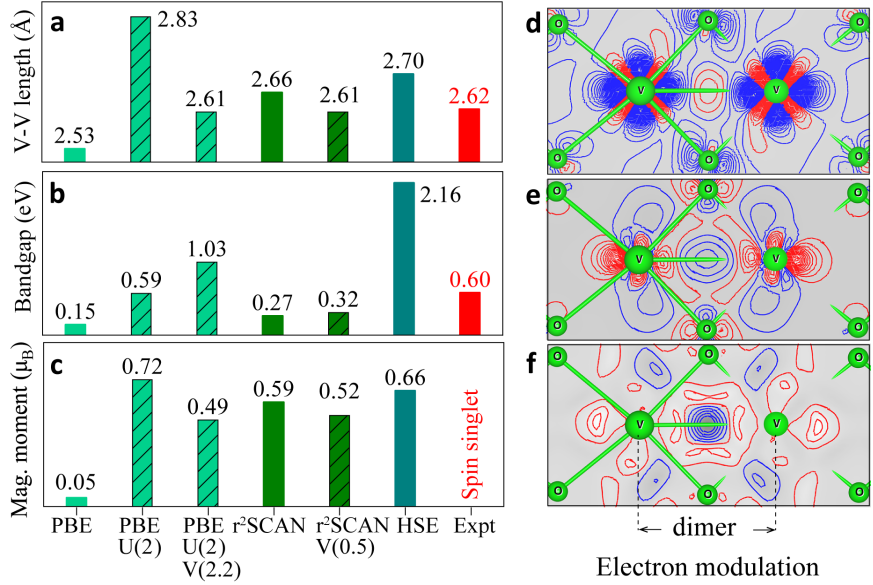

**Figure 3. Properties of VO$_2$.** (a) Length of vanadium-vanadium dimer. Note that the lattice constants are fixed to their experimental values due to technical constraints, and only the ionic positions are optimized. The experimental value is 2.62 Å [51]. (b) Bandgap. (c) Local magnetic moments. (d) Redistribution of electron density, $\Delta n = n^{\text{PBE}+U(2\text{eV})} - n^{\text{PBE}}$. The +$U$ correction induces inhomogeneous electron redistribution, causing accumulation in non-bonded $t_2$ orbitals and depletion from directional V-O bonds, while overall resulting in electron accumulation at the vanadium site. (e) $\Delta n = n^{\text{PBE}+U(2\text{eV})+V(2.2\text{eV})} - n^{\text{PBE}}$. (f) $\Delta n = n^{\text{r}^2\text{SCAN}+V(0.5\text{eV})} - n^{\text{r}^2\text{SCAN}}$. Red and blue colors denote electron depletion and accumulation, respectively.

We find that the PBE-based methods have notable limitations. First, PBE+$U$ with a small $U$ value of ~0.1 eV can well reproduce the experimental bond length (Figure S7), potentially giving the impression that PBE is nearly accurate for describing VO$_2$. However, this outcome is coincidental in VO$_2$, similar to what is observed in the Cr$_2$ molecule. Second, the PBE+$U$+$V$ method, with parameters deliberately chosen to reproduce the dimer length, introduces an adverse consequence: the bandgap is significantly overestimated (1.03 eV; see Figure 3b), which is atypical for local and semilocal functionals.

In contrast, the r$^2$SCAN+$V$ method addresses both on-site and on-bond electron localization with simplified parameterization. Using a $V$ value of 0.5 eV, r$^2$SCAN+$V$ accurately reproduces the bond length (Figure 3a) while yielding a reasonable bandgap (Figure 3b). The mechanism is again revealed from the electron accumulation on the nonlocal region (Figure 3f).

While the above simulations have adopted a spin-symmetry-breaking (SSB) treatment, VO$_2$ has a spin-singlet ground state for the vanadium dimer. SSB has been criticized [52] due to the emergence of local magnetic moments (Figure 3c). We discuss three advantages of the SSB treatment. First, several authors of this work has established that SSB is, in general, far more revealing than its spin-restricting counterpart [49,50,53]. SSB can reveal dynamic but slow spin fluctuations. Second, the SSB treatment on VO$_2$, which leads to electron deficiency on the bond, becomes physically accurate when combined with a positive intersite $V$, as it mitigates the deficiency by redistributing electrons back to the bond. In contrast, the spin-restricting treatment [52], which already causes electron over-accumulation on the bond, worsens the problem when combined with the corrective $V$ method. For instance, PBE+$V$ ($V$ = 2 eV) under spin-restricting treatment [52] amplifies the over-binding issues, resulting in an excessively short bond length of 2.46 Å (the lattice and ionic positions are simultaneously optimized) and an over-corrected bandgap of 0.62 eV. Finally, the broken symmetry introduced by SSB can be restored afterward [54], ensuring the correct wavefunction symmetry.

# 5. Fe with *d-d* covalent bonding and hidden antiferromagnetism

SCAN's applicability to transition-metal materials was initially questioned in the elemental Fe, revealing an overestimation of magnetic moment [27,55,56]. This issue invites comparisons with graphene, $Cr_2$, and $VO_2$. Here, we reveal that Fe also demonstrates non-compact covalent bonding, which occurs only between certain parts of the atomic shells and is concealed within a metallic background. To provide a comprehensive analysis, we systematically examine the bonding characteristics of four elemental metals—Cr, Fe, Co, and Ni—excluding Mn due to its large unit cell, using the projected Crystal Orbital Hamilton Populations (pCOHP) method [57,58]. As shown in Figure 4a, the valence bands of Cr are predominantly characterized by bonding states, with an increasing incorporation of antibonding states as we progress to Fe, Co, and Ni. This trend is quantitatively captured by the integrated pCOHP (Figure 4b), where the monotonically decreasing values reflect progressively weaker covalent bonding among the four materials.

We treat Cr as a collinear AFM material, not as a spiral static spin-density wave. As a rare case of an antiferromagnetic metal, Cr exhibits the most significant covalency. Fe also demonstrates considerable covalency and features an antiferromagnetic coupling between its $t_{2g}$ orbitals [59], concealed within a ferromagnetic background. The covalency and antiferromagnetism in Cr and Fe, associated with direct overlap of adjacent *d* orbitals, may suggest underlying physics similar to that we observed in graphene, the $Cr_2$ molecule, and $VO_2$. Indeed, the electron redistribution provides straightforward evidence: an inter-site *V* potential accumulates electrons on the shortest bonds in Cr and Fe (Figures 4c and 4d), but the pattern is absent in Co and Ni (Figure S9).

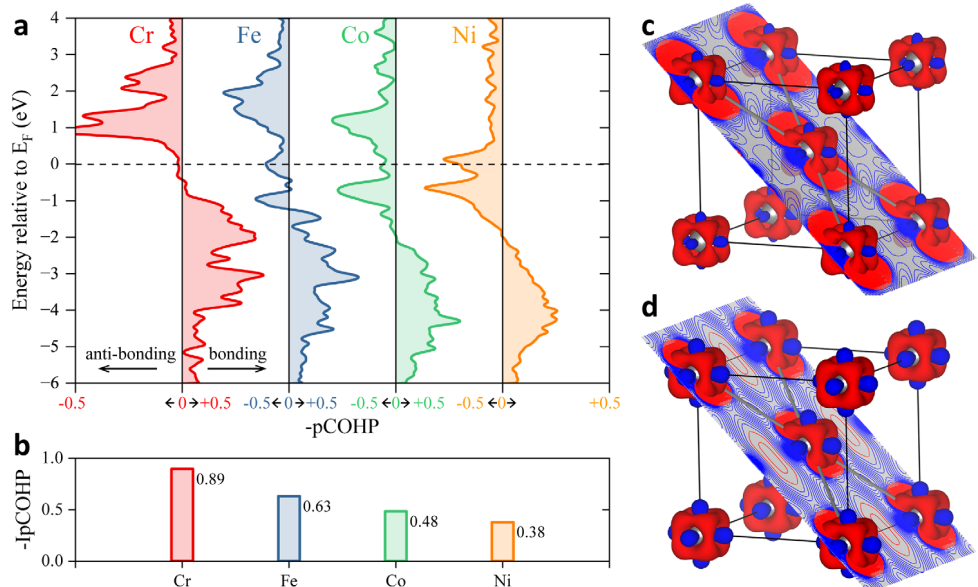

**Figure 4.** Bonding properties of transition metals. (a) The negative values of *projected Crystal orbital Hamilton populations*, –pCOHP. The PBE functional is used here; $r^2$SCAN results are presented in Figure S8. (b) Integrated value, –IpCOHP. (c) Electron redistribution in Cr, $\Delta n = n^{r^2SCAN+V(2eV)} - n^{r^2SCAN}$. (d) Electron redistribution for Fe, $\Delta n = n^{r^2SCAN+V(4eV)} - n^{r^2SCAN}$. The *V* potential is on the 3*d* orbitals. Red and blue isosurfaces/lines denote electron depletion and accumulation, respectively.

Having established the covalency feature in Cr and Fe, we calculate their magnetic moments using the $r^2$SCAN+*V* method with varying *V* values, as shown in Figure 5a. Without the +*V* correction, $r^2$SCAN alone predicts a magnetic moment of 2.12 $\mu_B$ for Cr, significantly higher than the experimental value of 0.51 $\mu_B$ [60]. For Fe, the result is similarly overestimated at 2.95 $\mu_B$ compared to the experimental result of 2.22 $\mu_B$ [61]. The +*V* correction effectively reduces the magnetic moments, with $V \approx 2.0$ eV for Cr and $V \approx 4.0$ eV for Fe, yielding results that closely match experimental values. In contrast, the +$V$ correction for Co and Ni has a negligible effect or slightly increases the magnetic moments, suggesting that this correction may not be applicable. The small overestimation of Ni's moment by r$^2$SCAN likely originates from other factors and warrants further investigation.

The optimal $V$ values in r$^2$SCAN+$V$, which are chosen to best reproduce specific experimental quantities, show considerable variation across the tested materials. For example, $V_{pp} \geq 2$ eV for graphene, $V_{dd} = V_{ss} = 0.8$ eV for the Cr$_2$ molecule, $V_{dd} = 0.5$ eV for VO$_2$, $V_{dd} = 2$ eV for Cr, and $V_{dd} = 4$ eV for Fe. Overall, the $V$ values are relatively small (except for Fe). These variations are driven by different electronic screenings of these materials, necessitating an advanced DFT functional capable of accurately capturing the subtlety. As a demonstration, Figure 5b compares the performance of various functionals, roughly organized by increasing magnetic moments. Among these, the OFR2 functional [46], designed to better account for electronic screening compared to r$^2$SCAN, predicts smaller magnetic moments. Thus, the magnetic moments serve as an informative indicator of how effectively DFT functionals describe the non-compact bonds.

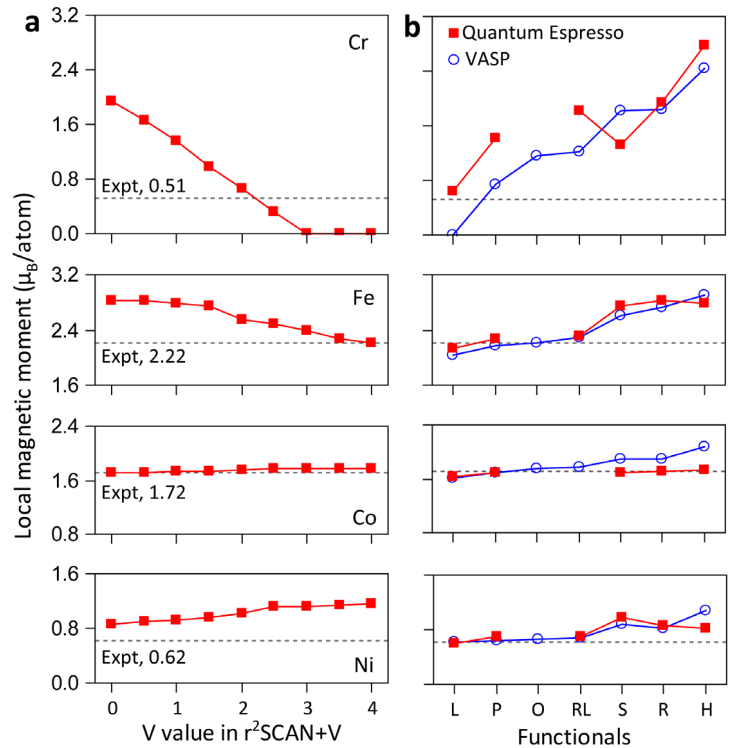

**Figure 5. Magnetic moment of transition metals.** (a) Simulation performed using r$^2$SCAN+$V$ with varying $V$ values. (b) Simulations conducted with various functionals, abbreviated as follows: **L** for LSDA, **P** for PBE, **O** for OFR2, **RL** for r$^2$SCAN-L, **S** for SCAN, **R** for r$^2$SCAN, and **H** for HSE06. VASP and Quantum Espresso are used for cross-checking, but the OFR2 functional is unavailable in Quantum Espresso. Experimental values for local magnetic moments are taken from Cr [60], Fe [61], Co [62], and Ni [61].

## 6. Discussion and Summary

We have proposed a possible way to understand the successes and puzzling failures of the non-empirical PBE and SCAN/r$^2$SCAN density functionals. Graphene, Cr$_2$ molecule, VO$_2$, Cr, and Fe exhibit notable non-compact covalent bonding through *s-s*, *p-p*, or *d-d* orbital hybridization, which can accumulate electrons on the bond centers. At the same time, the involved *s*, *p*, or *d* orbitals retain electrons around the atomic centers. It is crucial for DFT functionals to simultaneously account for electrons at local atomic sites and within non-compact bonds. PBE, while suffering from electron delocalization errors in both regions, accidentally strikes a balance in describing these materials (Figure 6a). In contrast, r$^2$SCAN enhances the localization of site-centered electrons more effectively than PBE, but it does not equally improve the localization of bond-centered electrons in non-compact bonds (Figure 6b). Such biased improvement compromises r$^2$SCAN's accuracy.

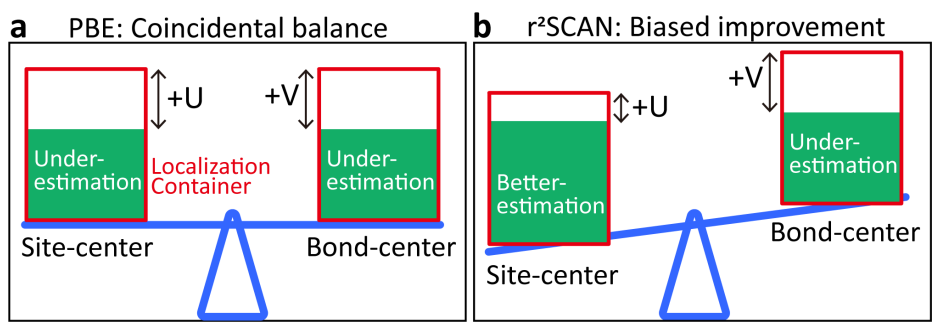

**Figure 6. Sketch of the site-centered and bond-centered electron localization.** (a) PBE's superior performance arises from coincidental error cancellation, despite pronounced delocalization errors in both site- and bond-centered regions. (b) r$^2$SCAN's underperformance results from its biased improvement at atomic sites.

The original GGA+$U$+$V$ method [31,32,33,34,35] could potentially enhance electron localization in both regions, yet it faces practical difficulties in determining $U$ and $V$ values in the tested materials. The r$^2$SCAN+$V$ approach simplifies parameter management and effectively resolves the accuracy challenges for all tested materials with typically small $V$ parameters. Future revisions of r$^2$SCAN may need to better account for this kind of inter-site effect. Whether this can be achieved within or only beyond the computationally-efficient meta-GGA functional form, in particular when including both kinetic energy density and Laplacian of electron density as its ingredients [63], and without material-dependent fitted parameters, remains to be seen. Fully non-local and thus less computationally-efficient self-interaction corrections [4,64] that make a functional exact for all one-electron densities require a unitary transformation to localized orbitals, including covalent-bond orbitals. A future refinement of Refs. [4,64] that does no harm to r$^2$SCAN accuracy (in the sense that LSIC-$\alpha$ [65] does no harm to LSDA accuracy) might boost the accuracy of r$^2$SCAN for the systems studied here, while satisfying an 18$^{th}$ exact constraint. Uncorrected SCAN and r$^2$SCAN are only approximately self-interaction-free, and only for compact one-electron densities [66,67].

Meta-GGA functionals, similar to SCAN or its modifications, are under active development. Recently, the Lebeda-Aschebrock-Kümmel (LAK) meta-GGA [68,69] demonstrated remarkable accuracy in predicting atomization energies, bond lengths, bandgaps, and capturing weak interactions near equilibrium. To address the over-magnetization issue of SCAN, a modification called mSCAN [70] "*provides a solution that satisfies the most pressing desiderata for density functional approximations in ferromagnetic, antiferromagnetic, and noncollinear states*". While these functionals offer significant improvements in many areas, their accuracy for certain cases, such as the Cr$_2$ molecule (Figure S5), still requires further refinement.

# Acknowledgments

YZ is supported by the Startup Grant of Minjiang University (30804326). TL and JS acknowledge the U.S. DOE, Office of Science, Basic Energy Sciences (BES), Grant No. DE-SC0014208. JPP was supported by the U.S. National Science Foundation under Grant No. DMR-2426275, and by the U.S. Department of Energy. Basic Energy Sciences, under Grant. No. DE-SC0018331. RM was supported by DMR-2426275.

# Conflict of Interest

The authors have no conflicts to disclose.

# Data availability

The data that support the findings of this study are available within the article and supplementary materials.